\documentstyle[floats,prl,aps]{revtex}

\newcommand{\bm}{\bbox}
\input epsf
\begin{document}
\draft
%%%%%%%%%%%%%%%%%%%%%%%%%%%%%%%%%%%%%%%%%%%%%%%%%%%%%%%%%%%%%%%%%%%%%%%%%%
%                               Title                                    %
%%%%%%%%%%%%%%%%%%%%%%%%%%%%%%%%%%%%%%%%%%%%%%%%%%%%%%%%%%%%%%%%%%%%%%%%%%
\twocolumn[\hsize\textwidth\columnwidth\hsize\csname @twocolumnfalse\endcsname
%\twocolumn[
\title{Van-Hove singularities effects on the upper critical field}
\author{R. G. Dias}
\address{Departamento de F\'{\i}sica, Universidade de Aveiro, \\
3810 Aveiro, Portugal.}
\date{\today}
\maketitle
%%%%%%%%%%%%%%%%%%%%%%%%%%%%%%%%%%%%%%%%%%%%%%%%%%%%%%%%%%%%%%%%%%%%%%%%%%
%                              abstract                                  %
%%%%%%%%%%%%%%%%%%%%%%%%%%%%%%%%%%%%%%%%%%%%%%%%%%%%%%%%%%%%%%%%%%%%%%%%%%
\widetext
\begin{abstract}
\leftskip 54.8pt \rightskip 54.8pt
We present a study of the
superconducting pairing susceptibility $K_T(r)$ for a
two-dimensional isotropic system with a strong power-law
divergence in the density of states $N(\epsilon) \sim
\epsilon^{-1+\frac{1}{b}}$, $b>1$. We show that the pair
propagator has the scaling form, $K_T(r)=r^{b-3} F(T^{1/b} r)$. An
anomalous short-range behavior is found,
leading straightforwardly to positive
curvature in the upper critical field, for $b\lesssim 2$
and to a zero temperature divergence,
$H_{c2} \sim T^{-2+\frac{4}{b}}$, for $b>2$.
\end{abstract}

]
%%%%%%%%%%%%%%%%%%%%%%%%%%%%%%%%%%%%%%%%%%%%%%%%%%%%%%%%%%%%%%%%%%%%%%%%%%
%                              body of paper                             %
%%%%%%%%%%%%%%%%%%%%%%%%%%%%%%%%%%%%%%%%%%%%%%%%%%%%%%%%%%%%%%%%%%%%%%%%%%
\narrowtext

Photoemission experiments in
Copper-Oxide superconductors \cite{gof}
have provided direct evidence for the existence of an extended
saddle point in the CuO$_2$ plane bands and consequently, a strong divergence
in the density of states,
$
      N(\epsilon) \sim (\epsilon-\epsilon_{vh} )^{-\alpha},
$
for energies close to $\epsilon_{vh}$.
Some authors have claimed that the high
superconducting critical temperatures of the cuprates
could be explained taking into account this divergence
\cite{lab,fri}.

One of the most surprising properties of these materials
is the upper critical field, which has been obtained
in magnetoresistance experiments
down to very low temperatures in the case of
overdoped Tl$_2$Ba$_2$CuO$_{6+\delta}$ \cite{mck}
and Bi$_2$Sr$_2$CuO$_y$ \cite{oso} and underdoped
YBa$_2$Cu$_3$O$_{7-\delta}$ \cite{wal}.
A very unusual $H_{c2}(T)$ curve is observed,
with very strong positive curvature and no evidence of saturation
at low temperatures.
This behaviour contrasts strongly with the  weak coupling
BCS result \cite{gor} which predicts an approximately
parabolic shape for the $H_{c2}$ curve.

Recently, Abrikosov has proposed \cite{abr1} that these anomalous
$H_{c2}$ curves
reflect a dimensional crossover to quasi-one dimensional
superconductivity due to the presence of flat regions
in the energy dispersion, that is, extended saddle points.
In Abrikosov's approach, the two extended saddle points in the 
energy dispersion are replaced by two one-dimensional linear 
energy dispersions $\epsilon_1(q_x)=v_1 \cdot q_x$ and 
$\epsilon_2(q_y)=v_2 \cdot q_y$. This model is equivalent to a 
system of two transverse chains and in this case, the density of 
states loses completely its strong energy dependence. Furthermore, 
it is not surprising that he finds a dimensional crossover in 
$H_{c2}$.
In this paper, we argue that these curves reflect not a
dimensional crossover, but the strong energy dependence
of the density of states which results from the presence
of these extended saddle points.
In the following, we present a
study of the superconducting pairing susceptibility
for a isotropic two-dimensional system with a
strong power-law divergence in the density of states.
As shown by Gorkov \cite{gor},
this two-particle correlation function determines the shape of
the superconducting transition $H_{c2}(T)$
of a type-II superconductor.

The superconducting
transition is characterized by the vanishing of
the gap function $ \Delta( {\bm{r}},{\bm{r^\prime}})$, defined as
$
     \Delta({\bm{r}},{\bm{r^\prime}})=
     V({\bm{r}}-{\bm{r^\prime}}) \langle\psi_\downarrow({\bm{r}})
     \psi_\uparrow({\bm{r^\prime}}) \rangle
$.
In the particular case of a local pairing interaction,
$V({\bm{r}}-{\bm{r^\prime}})=g\; \delta({\bm{r}}-
{\bm{r^\prime}})$, we obtain as usual the s-wave gap function,
$\Delta({\bm{r}},{\bm{r}}^\prime) =
\Delta({\bm{r}})\delta({\bm{r}}-{\bm{r}}^\prime)$.
In the following, $\hbar=c=e=k_B=1$.
In the vicinity of the
superconducting transition curve, the gap parameter is small
and a perturbation expansion in powers of $\Delta$ leads
to the semi-classical linearized gap equation \cite{gor,book}
\begin{equation}
         \Delta({\bm r}) = g\int d {\bm r^\prime}
         K_\beta({\bm r^\prime} -{\bm r })
         e^{i\,2{\bm A }({\bm r })  \cdot
         ({\bm r^\prime}-{\bm r })}
         \Delta({\bm r^\prime}).
         \label{eq:gapequation}
\end{equation}
where $K_\beta({\bm r})$ is the fermion pair propagator
in real space for a given temperature $T=1/\beta$,
in the absence of the external field and the pairing
interaction $g$ and is defined as
\begin{equation}
     K_\beta({\bm{r}}^\prime,{\bm{r}}) = {1 \over \beta} \sum_\omega
     {\cal G}_{-\omega}({\bm{r}}^\prime,{\bm{r}})
     {\cal G}_{\omega}({\bm{r}}^\prime,{\bm{r}}).
\end{equation}
where the Matsubara Green's function ${\cal G}_{\omega}$ describes the
normal state in the absence of magnetic field.
Using Kramers-Kronig relations,
$K_\beta({\bm r})$ can be rewritten  as
\begin{equation}
        K_\beta({\bm{r}})
        =  {2 \over \pi} \int d\omega
        \tanh(\beta \omega/2) A({\bm{r}},\omega) B({\bm{r}},-\omega),
        \label{eq:fouAB}
\end{equation}
with $A(\bm{k},\omega) = \mbox{Im} G^R(\bm{k},\omega)$
and $B(\bm{k},\omega) = \mbox{Re} G^R(\bm{k},\omega)$
where $G^R(\bm{k},\omega)$ is the retarded Green's function in the
absence of magnetic field and pairing potential.
$A(\bm{r},\omega)$
and $B(\bm{r},\omega)$ are the respective Fourier transforms.
A non-local $V({\bm{r}}-{\bm{r^\prime}})$
may lead to a d-wave gap solution and a slightly
modified gap equation. One can show that
the upper critical field probes the behaviour of
the Cooper pair center of mass and the internal symmetry of the
gap function is irrelevant as long as the thermal and magnetic
lengths are much larger than the interaction range.

In a bidimensional system, a van-Hove singularity (VHS) in the
density of states results usually from the presence of a saddle
point in the energy dispersion $\epsilon({\bf k})$.
In the case of an extended
saddle point, $\epsilon({\bm q}) \sim q_x^n - q_y^m$,
where ${\bm q}={\bm k}-{\bm k_{vh}}$,
this leads to a power-law divergence in  the density
of states $N(\epsilon)\sim \epsilon^{-1+{1\over n}+{1\over m}}$.
Such form for the extended saddle point is not only indicated by 
the direct probing of the energy dispersion using the angle 
resolved photoemission technique, \cite{gof} but also by numerical work on the 
Hubbard model. For instance, Quantum Monte Carlo work by 
Imada and collaborators \cite{imad} in the Hubbard model has found an extended saddle 
point with a quartic $q_y$ dependence at $(0,\pi)$.
The effect of such a divergence on the pairing susceptibility and
the superconducting phase diagram is the subject of this paper.
Clearly, a system with a saddle point is not isotropic. However,
in order to simplify the problem,
we adopt the isotropic dispersion relation:
\begin{equation}
     \epsilon({\bm k})-\epsilon_{vh}=a \cdot \text{sign}(q) \vert q\vert^b
     \label{eq:energy}
\end{equation}
where ${q}=k-k_{vh}$.
The influence of anisotropy in the semi-classical 
upper critical field is well 
studied. \cite{kle,leb}
Anisotropic two-dimensional systems have tipically 
open warped Fermi surfaces or elliptical closed Fermi surfaces
in the case of small particle number.
For a system with an elliptical Fermi surface, 
in the case of a transverse magnetic field, 
it is simple to show that the normalized upper critical 
field follows the parabolic-like BCS curve.\cite{thesis}
For an open warped Fermi surface, the behavior of the
$H_{c2}$ curve is determined by the relation 
between $T_{co}$ and 
the small $t_y$ modulation of the Fermi surface.\cite{thesis,leb}
If $t_y \ll T_{co}$, $H_{c2}$ will diverge at a finite 
temperature, reflecting a reduction of the effective dimension of 
our system induced by the magnetic field.\cite{kle,leb}
However, if $t_y \gg T_{co}$, a BCS-like parabolic $H_{c2}$
curve is obtained. A reduction of $t_y$ enhances the zero 
temperature critical field, $H_{co}$ relatively to the zero field critical
temperature, $T_{co}$,
but, as long as the relation is  valid, the reduced
upper critical field  ($H_{c2}/H_{co}$
as a function of $T/T_{co}$) remains unchanged.
One may therefore conclude that, unless a dimensional crossover
is present, the reduced upper critical field   
shows very little sensitivity to
anisotropy. Another important point is that the contribution
to superconductivity of the extended saddle-point region
is much larger than the contribution of the other regions of the 
Fermi surface, which can therefore be neglected.
These facts motivate the 
choice of an isotropic model for our study.

Note that, for quasi-2D systems and magnetic field applied along 
the planes, the dimensional crossover is from 3D to 2D. 
In the case of transverse fields, the crossover is from quasi-2D to 
quasi-1D superconductivity. 
One should note, however,  the $H_{c2}$
divergence results from a mean-field 
analysis and fluctuations modify this behavior greatly in the last case. 
In fact, saturation should 
arise at low temperatures due to fluctuations, reflecting the well known 
impossibility of a superconducting state in one dimension. 
Furthermore, as recently shown by Lebed et al \cite{leb}, saturation 
should also be observed due to Pauli pair breaking. 
Therefore, it seems unlikely that a dimensional crossover could explain 
results obtained by Mackenzie and others\cite{mck,oso}.

We assume that the VHS is pinned at the Fermi level, 
that is, $k_F=k_{vh}$. 
We will comment 
on the pinning assumption at the end of the paper.
The density of states for the above model is $N(\epsilon)\sim
a^{-1/b} b^{-1} (\epsilon-\epsilon_{vh} )^{{1\over b}-1}$.
Let us assume for now that $b$ is an odd integer.

For this simple model, we can compute the spectral function
\begin{eqnarray}
        A(r,\omega) &=& -\frac{1}{2ab}
        \left(\frac{\vert \omega \vert}{a}\right)^{{1\over b}-1}
        \cdot \sqrt{\frac{2k_F}{\pi r}} \nonumber \\
        & & \cos \left[r \left(\left(\frac{\vert \omega \vert}{a}\right)
        ^{1\over b} \text{sign} (\omega) +k_f \right) -\frac{\pi}{4}
        \right]
        \label{eq:spectral}
\end{eqnarray}
and the retarded Green´s function $G^R(r,\omega)$, since
$G^R(q,\omega)$ is a meromorphic function in the complex
$q$-plane. Note that
$A(r,\omega)= \mbox{Im} G^R(r,\omega)$ and $B(r,\omega) =
\mbox{Re}\; G^R(r,\omega)$. After some lengthy but straightforward algebra,
one obtains the following expression for the pair propagator,
\begin{equation}
        K_\beta(r)=
        r^{b-3}
        F\left[\left(\frac{\beta a}{2}\right)^{1\over b} / r\right]
        \label{eq:pairprop}
\end{equation}
with
\begin{eqnarray}
        F[X] &=& \frac{2k_F}{\pi^2} \frac{1}{ab}
        \int_0^\infty d\omega \frac{\tanh [(\omega X)^b]}
        {\omega^{b-1}} \left\{\frac{1}{2} \sin (2\omega) \right.   \nonumber\\
        &+&
        \left. \sum_{n=1}^{\frac{b-1}{2}} e^{-\omega \sin({2\pi \over b}n)}
        \sin\left[w(1+\cos({2\pi \over b}n))+{2\pi \over b}n\right]
        \right\}
.
        \label{eq:FX}
\end{eqnarray}
When $X\gg 1$, $F[X]\sim X^{b-2}$ and for $X\ll 1$, the
function is exponentially small. Note that the thermal length is
given by $\xi_T \sim (a/T)^{1/b}$. The pair propagator for
distances smaller that the thermal length is approximately given
by $K_T(r) \sim r^{-1} T^{-1+\frac{2}{b}}$ and therefore, it
diverges as the temperature goes to zero. We will show that this
will lead to a zero temperature divergence in the upper critical
field. Note that no Debye-like frequency cutoff was introduced in the
previous integral. This procedure is valid as long as the
temperature provides a smaller cutoff in the integrand, that is,
$T^{1/b} \ll \omega_c$. This reflects the well known reduction
of the isotope effect in the van-Hove scenario.\cite{isot}

The zero field critical temperature is obtained from the equation
\begin{equation}
        1/g=\frac{k_F}{\pi}\frac{1}{a^{1/b}b} \int_0^\infty d\omega
        \tanh (\beta \omega/2) \omega^{\frac{1}{b}-2}
\end{equation}
which leads to
\begin{equation}
        T_{co} \sim \left[\frac{k_F}{\pi}\frac{a^{-\frac{1}{b}}}{b-1}
        g\right]^{b/(b-1)}
\end{equation}
This result and the role of the frequency cutoff can be
qualitatively understood using the usual BCS relation for the
critical temperature $T_{co} \sim \omega_c e^{-1/ \langle N(\epsilon)
\rangle_{T_{co}} g}$, where $\langle N(\epsilon) \rangle_{T_{co}}$ is
the thermal averaged density of states $\langle N(\epsilon)
\rangle_{T} \sim \int d\epsilon (\partial f/\partial \epsilon)
N(\epsilon)$ and $f$ is the Fermi distribution function. In this
case, $\langle N(\epsilon) \rangle_{T_{co}} \sim a^{-1/b} b^{-1}
T_{co}^{{1\over b}-1}$ and therefore $\ln (\omega_c/T_{co}) T_{co}^{{1\over
b}-1} \sim 1/g$. In the weak coupling limit, one can neglect the
logarithmic correction and thus, the above dependence for the
critical temperature is reproduced. In the usual case of an
extended saddle point, this broadening argument leads correctly to
a transition temperature \cite{lab2,abr} $T_{co} \propto g^2$. The
enhancement of the critical temperature is clearly bounded by the
equivalent of the Debye Temperature in this problem, that is,
$T_{\text{lim}} \sim \omega_D$, where $\omega_D$ is our cutoff in
frequency. The energy dispersion as given by Eq.~\ref{eq:energy} may also be
limited to an energy range $\omega_c < \omega_D$ and in that case
$T_{\text{lim}} \sim \omega_D e^{-1/ \langle N(\epsilon)
\rangle_{\omega_c} g}$. This dependence on the extent of the
anomalous energy dispersion could offer an explanation for the low
critical temperatures of, for example, Bi2201 \cite{kin} and
Sr$_2$RuO$_4$ \cite{lu}. Photoemission experiments in these
materials \cite{kin,lu} have found a VHS near
the Fermi level, but also a smaller extent of the flat regions in
the energy dispersion. A similar thermal broadening
argument can be applied to the
zero temperature slope of the $H_{c2}$ curve.

%%%%%%%%%%%%%%%%%%%%%
%      figure       %
%%%%%%%%%%%%%%%%%%%%%
\begin{figure}[tbp]
       \begin{center}
       \leavevmode
       \hbox{%
       \epsfxsize 3.0in \epsfbox{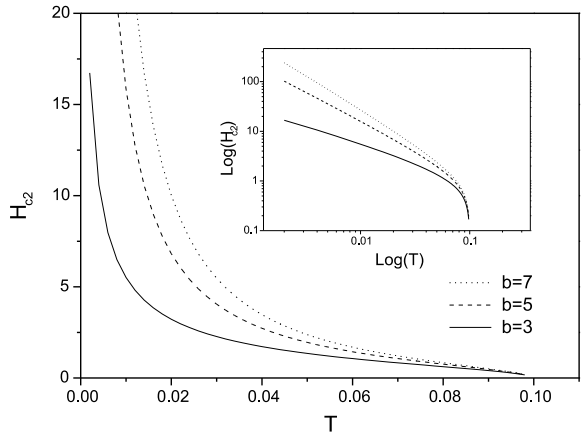}}
       \end{center}
       \caption{$H_{c2}$ curves for $b$=3, 5 and 7,
       obtained numerically from Eq.~\ref{eq:numeric}. Inset: The same curves
       in a logarithmic scale, showing clearly the low temperature
       power-law behavior.}
       \label{fig:hc2}
\end{figure}
%%%%%%%%%%%%%%%%%%%%%
%      end          %
%%%%%%%%%%%%%%%%%%%%%

\begin{figure}[tbp]
       \begin{center}
       \leavevmode
       \hbox{%
       \epsfxsize 3.0in \epsfbox{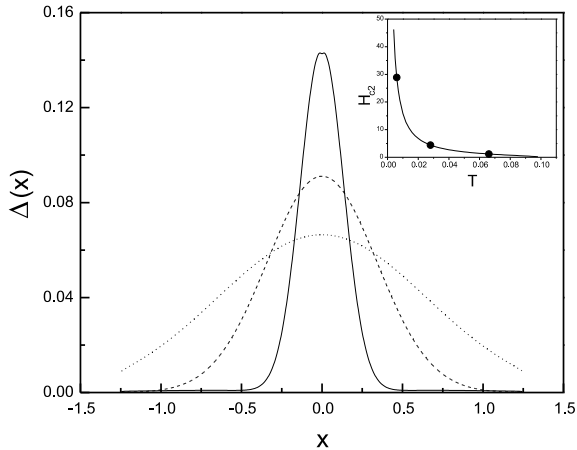}}
       \end{center}
       \caption{The numerically obtained gap solution at different points of the
       transition curve shown in the inset ($b=5$). All solutions are  Gaussian.}
       \label{fig:gapfunction}
\end{figure}

The analytical determination of the
upper critical curve for the complete temperature range is a difficult task.
So, we obtain the $H_{c2}$ curves by numerical solution of the
gap equation and study its behavior analytically only at low temperatures or
close to $T_c$. For numerical purposes, it is more
convenient to work with the gap function in a mixed representation.
If one chooses the Landau
gauge ${\bm A }=(0,Hx,0)$ and makes use of the degeneracy of the
gap function, one can rewrite
Eq.~\ref{eq:gapequation} as
\begin{equation}
         \widetilde{\Delta}(x) = g\int dx^\prime
         \widetilde{K}_\beta[x^\prime-x,-H(x+x^\prime)]
         \widetilde{\Delta}(x^\prime),
         \label{eq:numeric}
\end{equation}
where $\widetilde{\Delta}(x)$ is the $y$ integrated gap function and
$\widetilde{K}_\beta (x,k_y)$ is the Fourier transform of $K_\beta (x,y)$.
At zero temperature, the gap equation simplifies to
\begin{equation}
         \Delta({\bm r}) \sim  g\int d {\bm r^\prime}
         {T^{1-{2 \over b}} \over \vert r^\prime -r \vert }
         e^{i \phi({{\bm r^\prime} \over \sqrt{H}}, {{\bm r } \over
         \sqrt{H}})}
         \Delta({\bm r^\prime}).
\end{equation}
where $\phi$ is the magnetic phase acquired by the Cooper pair, which is independent
of the magnetic field if $r$ is written in magnetic length units.
Note that this form for the gap equation is independent of our gauge choice.
The numerical gap solutions show a perfect scaling $\widetilde{\Delta}(x)=
F(x/\sqrt{H})$, that is, all gap solutions fall into an universal Gaussian
curve (see Fig.~\ref{fig:gapfunction}), if the x-axis unit is the magnetic length
and therefore, with the variable change
$\widetilde{x}= x/\sqrt{H}$ in the previous equation,
the gap function becomes independent of $H$
and  we obtain the low temperature scaling of the upper critical field,
$H_{c2}(T) \sim T^{-2+{4 \over b}}$.

In Fig.~\ref{fig:hc2}, $H_{c2}$ curves for several values of $b$,
obtained numerically from Eq.~\ref{eq:numeric} are displayed.
These curves are characterized by a strong divergence of the upper critical
field as $T \rightarrow 0$ and linear behavior close to
$T_{co}$. In the inset, the low temperature scaling is
clearly observed
in a log-log scale.
This behavior is clearly distinct from a dimensional crossover in
$H_{c2}$ which would lead to a divergence even in
a log-log scale.

While Eq.~\ref{eq:pairprop} for the pair propagator
was derived for odd integer
$b$,
we believe this equation is qualitatively correct for any value of $b \geq 1$.
Clearly, Eq.~\ref{eq:spectral} for the spectral function is valid for any $b$
and one
 can show that the pair propagator for $b > 1$ will have the same qualitative
short
and long range behavior as that given by Eq.~\ref{eq:pairprop}.
For $b=1$, with the introduction of a cutoff,
we recover the usual BCS results and, in particular,
$H_{co} \sim T^{2}_{co}$. For $1\leq b< 2$, $F[X]\sim \text{const}$, if $X\ll 1$
and therefore, the pair propagator shows a different short range
dependence, $K_T(r) \sim r^{b-3}$.
%%%%%%%%%%%%%%%%%%%%%
%      figure       %
%%%%%%%%%%%%%%%%%%%%%
%%%%%%%%%%%%%%%%%%%%%
%      end          %
%%%%%%%%%%%%%%%%%%%%%
The pair propagator does not diverge
as we decrease the temperature, and with a scaling argument, we can show
that now the zero temperature critical field is finite,
$1/g \sim H^{(1-b)/2}_{co}$ and thus, $H_{co} \sim T^{2/b}_{co}$.

The low temperature dependence of $H_{c2}$ can be obtained expanding the
pair propagator in powers of $T$,
\begin{equation}
        [K_T(r)-K_0(r)]/ r^{b-3} \sim -(rT^{1/b})^c
\end{equation}
and following Gorkov \cite{gor},
one obtains $H_{c2}(T)-H_{c2}(0)\sim -T^{2c/b}$.
Curiously, a power-law low temperature dependence of $H_{c2}$ has
also been suggested by Kotliar and Varma \cite{var} as a consequence of a zero
temperature critical point.
This dependence, in our picture, results
from the scaling form of the pair propagator as given by Eq.~\ref{eq:pairprop},
but the value of $c$ depends on the specific form of the integrand of Eq.~\ref{eq:FX}.
One knows that when $b \rightarrow 1$, the usual expression for the
pair propagator should be recovered, which is the one given by Eqs.~\ref{eq:pairprop} and
\ref{eq:FX} only with the sine function in the integrand \cite{book,rdias}.
For $b>2$, the exponential
term in Eq.~\ref{eq:FX} dominates and the sine contribution becomes irrelevant. Therefore,
when $b \rightarrow 1$, the low temperature behavior should be determined by the sine term
and as $b$ goes away from 1, the exponential term should take over. With this assumption,
$c$ can be determined and the result is
$c = (2-b)/2$, when $b \sim 2$ and $c = (3-b)/2$, when $b \rightarrow 1$.

%%%%%%%%%%%%%%%%%%%%%
%      figure       %
%%%%%%%%%%%%%%%%%%%%%
\begin{figure}[tbp]
       \begin{center}
       \leavevmode
       \hbox{%
       \epsfxsize 3.0in   \epsfbox{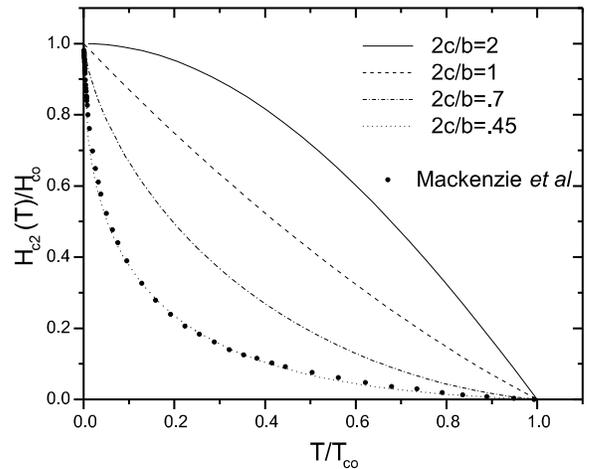}}
       \end{center}
       \caption{The qualitative normalized $H_{c2}$ curves for $1<b<2$.
        As the dispersion
        relation changes from linear to quadratic, the upper critical curve
        changes from the usual BCS curve to a curve with strong
        positive curvature. }

       \label{fig:qual}
\end{figure}
%%%%%%%%%%%%%%%%%%%%%
%      end          %
%%%%%%%%%%%%%%%%%%%%%

The results obtained up to now can be collected into a equation
similar to the usual one,\cite{rdias} $1/g=\int dr K_\beta(r) e^{-r^2 H}$
with a qualitative pair susceptibility given by
\begin{equation}
        K_T(r)={1 \over r^{3-b}}  { \left({r T^{1/b}}
        \right)^c \over
        \sinh \left[\left({r T^{1/b}}\right)^d \right]}
\end{equation}
with $c=0$ and $d=b-2$, if $b>2$. If $1\leq b<2$,
$c=d$ with $c$ having the behavior described above
in order to reproduce the low temperature dependence of the upper critical field.
In particular, $c=1$, if $b=1$ and the usual
BCS equation is recovered \cite{book,rdias}. If $b \rightarrow 2$, $c \rightarrow 0$.
In Fig.~\ref{fig:qual}, $H_{c2}$ curves obtained with this qualitative kernel
are displayed. A drastic transformation from conventional parabolic-like
curves (obtained with $c = (3-b)/2$)
to  curves with strong positive curvature (obtained with $c = (2-b)/2$)
is observed as the low temperature exponent $2c/b$ goes from 2 to 0.

In Fig.~\ref{fig:qual}, the experimental $H_{c2}$ points obtained
by  Mackenzie {\it et al} for Tl$_2$Ba$_2$CuO$_{6+\delta}$
\cite{mck} are also displayed and fitted with our qualitative
$H_{c2}$ curves. Note that this is a one-parameter fit ($c =
(2-b)/2$), since the normalized curves do not depend on the
coupling constant $g$. An impressive agreement is observed for
$2c/b=.45$, which according to the picture presented in this
paper, implies that the density of states diverges as $
N(\epsilon) \sim \epsilon^{-.28}$. 
Most photoemission experiments have found saddle points with quadratic 
dispersion along one direction and much flatter (higher 
power dependence) behavior along the other (transversal) direction, 
indicating therefore a divergence exponent $\alpha$ smaller than 1/2. 
In the case of saddle point obtained in \cite{abr}, 
a good fit is obtained with a quartic dependence, 
leading to $\alpha \approx 0.25$ which 
agrees reasonably with the value extracted from the experimental $H_{c2}$ 
curve.
We emphasize that for a given
exponent $\alpha$, the normalized $H_{c2}$ curve is as universal
as the usual BCS curve \cite{gor} ($\alpha=0$). A suggestion of
some sort of universality is indeed observed in Fig.~4 of
Ref.~[6], where $H_{c2}$ curves of two different materials,
YBa$_2$(Cu$_{0.97}$Zn$_{0.03}$)$_3$O$_{7-\delta}$ and
Tl$_2$Ba$_2$CuO$_{6+\delta}$, apparently fall onto the same curve
in a plot of reduced $H_{c2}(T)$ versus reduced temperature. Such
universal $H_{c2}$ behavior is not observed in the case of a
simple saddle point which leads to a weak logarithmic divergence
in the density of states. \cite{rdias2}
 In this case, $H_{c2}$
depends on the coupling constant $g$ and shows upward curvature 
which becomes stronger as $g$ is decreased. Note that 
$H_{co}$ and $T_{co}$ have a weaker enhancement in this case,\cite{rdias2} 
with  $\log$ squared  relations to the inverse of the coupling constant
$g$, while for the extended singularity  power-law relations
have been obtained. 

%%%%%%%%%%%%%%%%%%%%%%%%%%%%%%%%%%%%%%%%%%%%%%%%%%%%%%%%%%%%%%%%%%%%%%%
%                             Discussion                              %
%%%%%%%%%%%%%%%%%%%%%%%%%%%%%%%%%%%%%%%%%%%%%%%%%%%%%%%%%%%%%%%%%%%%%%%

It has been assumed throughout the paper that the VHS was pinned 
at the Fermi level. 
It has been shown in Ref.\cite{rdias2} that a deviation from the Fermi level
of the VHS leads to $H_{c2}$ saturation at low temperature,
the temperature range of this region being proportional to the energy difference,
$k_B T_{cross} \sim E_F-E_{vh}$. However, this has not been observed in the
experimental $H_{c2}$ curves,\cite{mck,wal} even though the respective samples
are in the overdoped or underdoped regime. According to the van-Hove scenario,
one should expect a certain deviation of the VHS from the Fermi level in these
regimes in order to account 
for the decrease of the zero field critical temperature.
However, it is possible that, (at least, in some interval of doping range)
the reduction of critical
temperature  results not from the deviation of the VHS from the Fermi level,
but, instead, from the weakening of the VHS due to a reduction of the
extent of the saddle point as suggested by King {\it et al}. \cite{kin} 
Photoemission experiments on 
YBa$_2$Cu$_3$O$_{6.9}$, \cite{cam}
YBa$_2$Cu$_3$O$_{6.5}$, and
YBa$_2$Cu$_3$O$_{6.3}$ \cite{liu}
support this picture, since they report
a clear doping independence of the pinning of the  Fermi level
at the VHS.
Moreover, this doping independence of the pinning is predicted by many
numerical studies, from slave-boson calculations \cite{mar,new}
to renormalization group calculations \cite{gon}.

In conclusion, we have studied the effect of a power-law
divergence of the density of states at the Fermi level
$N(\epsilon) \sim \epsilon^{-\alpha}$ on the upper critical field
of a clean isotropic weak-coupling superconductor. We have shown
that for a weak divergence ($\alpha$ less than 1/2), the zero
temperature critical field is finite, but strong positive
curvature appears in $H_{c2}$ as $\alpha$ approaches 1/2. For a
stronger divergence ($\alpha$ larger than 1/2), $H_{c2}(T)$ has a
power-law divergence at $T=0$. A very good one-parameter fit was
obtained to the experimental results by Mackenzie {\it et al}
\cite{mck}. According to the picture described in this paper, the
anomalous $H_{c2}$ behavior reflects the short-range enhancement
of the pair propagator and the unusual temperature dependence of
the thermal length which result from the existence of a strong
divergence of the density of states at the Fermi level.

%%%%%%%%%%%%%%%%%%%%%%%%%%%%%%%%%%%%%%%%%%%%%%%%%%%%%%%%%%%%%%%%%%%%%%%
%                            acknowledgements                         %
%%%%%%%%%%%%%%%%%%%%%%%%%%%%%%%%%%%%%%%%%%%%%%%%%%%%%%%%%%%%%%%%%%%%%%%
This research was funded by the Portuguese \linebreak
MCT PRAXIS XXI program under Grant No. 2/2.1/Fis/302/94.

%%%%%%%%%%%%%%%%%%%%%%%%%%%%%%%%%%%%%%%%%%%%%%%%%%%%%%%%%%%%%%%%%%%%%%%
%                                                                     %
%%%%%%%%%%%%%%%%%%%%%%%%%%%%%%%%%%%%%%%%%%%%%%%%%%%%%%%%%%%%%%%%%%%%%%%

\end{document}